\documentclass{ws-ijmpd}
\usepackage[super,compress]{cite}
\begin{document}

\markboth{R.Kh. Karimov, R.N. Izmailov and K.K. Nandi}
{Terrestrial Sagnac delay in scalar-tensor-vector-gravity}

\catchline{}{}{}{}{}
%
%%%%%%%%%%%%%%%%%%%%%%%%%%%%%%%%%%%%%%%%%%%%%%%%%%%%%%%%%%%%%%%%%%%%

\title{Terrestrial Sagnac delay in scalar-tensor-vector-gravity}

\author{R.Kh. Karimov$^{\ast}$}

\address{Zel'dovich International Center for Astrophysics, M. Akmullah Bashkir State Pedagogical University, 3A, October Revolution Street, Ufa 450008, RB, Russia \\
$^{\ast}$karimov\_ramis\_92@mail.ru}

\author{R.N. Izmailov$^{\dag}$}

\address{Zel'dovich International Center for Astrophysics, M. Akmullah Bashkir State Pedagogical University, 3A, October Revolution Street, Ufa 450008, RB, Russia, \\
Institute of Molecule and Crystal Physics, Ufa Federal Research Centre, Russian Academy of Sciences, Prospekt Oktyabrya 151, Ufa 450075, Russia \\
$^{\dag}$izmailov.ramil@gmail.com}

\author{K.K. Nandi}

\address{Zel'dovich International Center for Astrophysics, M. Akmullah Bashkir State Pedagogical University, 3A, October Revolution Street, Ufa 450008, RB, Russia, \\
High Energy Cosmic Ray Research Center, University of North Bengal, Darjeeling 734 013, WB, India \\
kamalnandi1952@rediffmail.com}

\maketitle

\begin{history}
\received{Day Month Year}
\revised{Day Month Year}
\end{history}

\begin{abstract}

\end{abstract}


\ccode{PACS numbers:}

%\tableofcontents
\section{Abstract}

The scalar-tensor-vector-gravity (STVG), a prototype of modified gravity developed by Moffat, can correctly explain galaxy rotation curves, cluster dynamics, Bullet Cluster phenomena and cosmological data without invoking the observationally elusive general relativistic (GR) dark matter. Further, recent observations of neutron star masses are shown to defy some GR predictions, whereas STVG turns out to be more consistent with those observations. These successes indicate that STVG could be a potential candidate for a new theory of gravity. However, an important question concerns the possible range of values of the STVG dimensionless parameter $\alpha$ imposed by various physical scenarios. In the literature, the range $0.03<\alpha <2.47$ corresponding to different central source masses has been suggested. We show here that the $\alpha$ can be considerably constrained into the range $0<\alpha<10^{-5}$ assuming that the updated GPS fluctuation does not exceed the $\alpha$-dependent correction to the terrestrial Sagnac delay.

\section{Introduction}

The general relativity (GR) postulate of attractive dark matter and repulsive dark energy, invoked to explain respectively the galactic flat rotation curves and cosmological acceleration, continues to remain enigmatic. Despite several speculations (WIMPs, scalar fields, etc), all experiments to date have failed to directly detect the elusive dark matter \cite{Aprileetal:2012,Agneseetal:2014,Akeribetal:2016,Akeribetal:2017}. This situation has led physicists to devise alternative theories, classified under the name MOdified Gravityy (MOG) models that seek to preserve the successes of GR but do not require the postulate of undetectable dark matter. Some notable MOG are, but not limited to, Weyl conformal gravity \cite{MO:2011} (see also Ref.~\refcite{NB:2012}), MOdified Newtonian Dynamics (MOND) \cite{Milgrom:1983}, $f(R)$-gravity \cite{NO:2006} and TeVeS \cite{Bekenstein:2004}.

Moffat \cite{Moffat:2006} developed an MOG, called the relativistic scalar-tensor-vector-gravity (STVG), which is based on a gravitational coupling constant $G$ that is assumed to exceed Newton's constant $G_{N}$ as $G=G_{N}(1+\alpha)$, $\alpha>0$. This assumption together with the introduction of a gravitational Yukawa-like repusive acceleration can successfully explain galactic flat rotation curves without the need for dark matter \cite{Moffat:2006,MR:2013}.

Before proceeding further, certain things need to be clarified. With the identification $Q=\sqrt{\alpha G_{N}}M$, the Kerr-STVG solution reduces to that of the Kerr-GR at $\alpha=0$ or in the static case ($a=0$), to Reissner-Nordstr\"{o}m of GR, as noted in the Ref.~\refcite{NB:2012}. However, the \textit{physical content} of the solution is quite different from that of GR as the quantity $Q$ is not the independently prescribable Coulomb charge as allowed in GR. On the contrary, it is called the gravitational charge $Q$ since it is determined solely by the mass itself. The effect of this $Q$ is symbolized by a repulsive gravitational Yukawa-like term appearing in the radial acceleration [see Eq.(20) below]. Therefore, the results of Kerr-STVG in our paper do \textit{not} belong to those of Kerr-GR, unless $\alpha=0$.

Apart from the local Schwarzschild tests of gravity, the STVG explains the dynamics of globular clusters in the galactic halo \cite{BM:2006,BM:2007,MT:2008}, and cosmological observations \cite{MT:2013}. Properties of accretion disks around in STVG and a detectable influence $\alpha$ on the accretion characteristics of supermassive black hole (BH) has been studied \cite{PLR:2017}. Stability of circular orbits around a BH in the ambience of a weak magnetic field has been studied \cite{HJ:2015}. The latest success of MOG \cite{MT:2019} is that it can reproduce with reasonable accuracy the observed velocity dispersion in the ultra-diffuse galaxy $NGC1052-DF2$ that shows it to be devoid of dark matter \cite{vanDokkumetal:2018}. Moffat and his collaborators studied the STVG theory in the generic context of cluster dynamics \cite{MR:2014} with successful application to the Bullet Cluster \cite{BM:2007}. Recently this theory has also been considered in the context of neutron stars \cite{LR:2017} and recent observation of gravitational waves \cite{Moffat:2016}. In all these cases, the observations are claimed to be in good agreement with the theory. Gravitational lensing \cite{LS:2013,Moffat:2015a,Nandietal:2017,MRT:2018,Lukmanovaetal:2018,Nandietal:2018,Izmailovetal:2019,Izmailovetal:2019aa,IL:2020,Izmailovetal:2020aa} and relative time delay \cite{Izmailovetal:2019bb,Izmailovetal:2020bb,Tuleganovaetal:2020} are another useful diagnostics that are able to distinguish the predictions from competing theories or a type of the objects. The intriguing connection of STVG BH with that of the brane-world BH has been pointed out and discussed at length in the context of weak and strong field gravitational lensing observables \cite{Izmailovetal:2019}.

Next we briefly describe the Sagnac effect under consideration in this paper. Consider a circular turntable of radius $R$ having a light source/receiver (meaning the source \textit{and} the receiver at the same point) fixed to the turntable. A beam of light split into two at the source/receiver are made to follow the same closed path near the rim in opposite directions before they are re-united at the source/receiver. In the limit, if the turntable is not rotating, the beams will arrive at the same time at the source/receiver and an interference fringe will appear. When the turntable rotates with angular velocity $\omega_{0}$, the arrival times at the source/receiver will be different for co-rotating and counter-rotating beams: longer in the former case and shorter in the latter. This difference in arrival times is called the Sagnac delay, which is measured by superimposing the two arriving beams with phase differences causing a shift in the interference fringes. The delay or fringe shift is a consequence of the lack of simultaneity (asynchrony) for motion of light signals along a closed loop.

The total arrival time lag between the two light beams, as measured at the source/receiver, can be obtained from special relativity, which gives, to first order in $\omega_{0}$,
\begin{equation}
\delta\tau_{S} = \frac{4\mathbf{\omega}_{0}\mathbf{S}}{c^{2}},
\end{equation}%
where $\mathbf{S}$ is the vector area of the projection, orthogonal to the rotation axis, of the closed path followed by the waves contouring the turntable, $c$ is the speed of light in vacuum and $\mathbf{\omega}_{0}$ is the angular velocity of the turntable. One remarkable modern use of the Sagnac effect lies in the global navigational systems in which the rotation of Earth needs to be taken into account while using radio signals to synchronize clocks. This advantage was also used by Allan, Weiss and Ashby \cite{AWA:1985} to arrive at more precise delay measurements. The effect has been previously worked out in different solutions of gravity, e.g., in the Kerr-Sen string metric \cite{BNN:2002}, in the Johansen-Psaltis metric \cite{KDBK:2019}, in the Brans-Dicke theory \cite{Nandietal:2001,Nandietal:2008}. However, the earliest work on Sagnac delay in GR, to our knowledge, was done by Ashtekar and Magnon \cite{AM:1975}. Tartaglia \cite{Tartaglia:1998} calculated the corrections to the delay due to mass and spin in the Kerr metric, and here we shall follow his methodology. There have been several recent works on the possibilities of measuring the Sagnac effect caused by a spinning gravitating source \cite{Tartaglia:2018,DiVirgilio:2010,Obukhov:2016}.

In this paper, we shall calculate corrections to the Sagnac delay (1) due to mass $M$, spin $a$ and the dimensionless STVG parameter $\alpha$, when the "turntable" is a massive spinning object, the Earth, and motions of the source/receiver (clock) around the Earth. We shall apply our obtained formulas to the one round-trip terrestrial Sagnac delay using the observed uncertainty in the updated GPS data to constrain $\alpha$. The constraint is only an \textit{upper} bound, so nothing restricts $\alpha$ to be actually still smaller arbitrarily approaching zero depending on the accuracy of measurement.

The paper is organized as follows. In Sec.2, we revisit the STVG theory and its Kerr-MOG BH solution and in Secs.3 and 4, we consider the round trip non-geodesic and geodesic equatorial orbits. In Sec.5, we calculate upper limits on $\alpha$ from the terrestrial Sagnac delay for non-geodesic motion (the Hafele-Keating lack of time synchrony) and in Sec.6, we shall use the GPS update for geodesic motion to propose a more severe constraint on $\alpha$. Sec.7 concludes the paper. We take $c=1$ unless specifically restored.

\section{STVG theory and the Kerr-MOG BH}

(a) We shall briefly revisit the STVG theory for explaining the involved terms and the underlying assumptions pointing out exactly where STVG differs from Einstein-Maxwell theory. The phenomenological action in the STVG theory is \cite{Moffat:2006}

\begin{eqnarray}
S &=& S_{\text{GR}} + S_{\phi} + S_{\text{S}} + S_{\text{M}}, \\
S_{\text{GR}} &=& \frac{1}{16\pi}\int d^{4}x\sqrt{-g}\frac{1}{G}R, \\
S_{\phi} &=& -\int d^{4}x\sqrt{-g}\left(\frac{1}{4}B^{\mu\nu}B_{\mu\nu} - \frac{1}{2}\widetilde{\mu}^{2}\phi^{\mu}\phi_{\mu}\right), \\
S_{\text{S}} &=& \int d^{4}x\sqrt{-g}\frac{1}{G^{3}}\left[\frac{1}{2}g^{\mu\nu}\nabla_{\mu}G\nabla_{\nu}G - V(G)\right] \nonumber \\
&& +\int d^{4}x\frac{1}{\widetilde{\mu}^{2}G}\left[\frac{1}{2}g^{\mu v}\nabla_{\mu}\widetilde{\mu}\nabla_{\nu}\widetilde{\mu} - V(\widetilde{\mu})\right].
\end{eqnarray}
Here, $g_{\mu\nu}$ is the spacetime metric, $R$ denotes the Ricci scalar, and $\nabla_{\mu}$ is the covariant derivative, $\phi^{\mu}$ stands for a
Proca-type massive vector field with mass $m_{\phi}$:

\begin{equation}
B_{\mu\nu} = \partial_{\mu}\phi_{\nu} - \partial_{\nu}\phi_{\mu},
\end{equation}%
$G(x)$ and $\widetilde{\mu }(x)$ are dynamical scalar fields that vary in space and time, $V(G)$, and $V(\widetilde{\mu })$ are the corresponding potentials. The full energy-momentum tensor is

\begin{eqnarray}
T_{\mu\nu} &=& T_{\mu\nu}^{\text{M}} + T_{\mu\nu}^{\phi} + T_{\mu\nu}^{\text{S}}, \\
T_{\mu\nu}^{\text{M}} &=& -\frac{2}{\sqrt{-g}}\frac{\delta S_{\text{M}}}{\delta g^{\mu\nu}}, \\
T_{\mu\nu}^{\phi} &=& -\frac{2}{\sqrt{-g}}\frac{\delta S_{\phi}}{\delta g^{\mu\nu}}, \\
T_{\mu\nu}^{\text{S}} &=& -\frac{2}{\sqrt{-g}}\frac{\delta S_{\text{S}}}{\delta g^{\mu\nu}}.
\end{eqnarray}

In the above, $T_{\mu\nu}^{\text{M}}$ denotes the ordinary matter energy-momentum tensor, $T_{\mu\nu}^{\phi}$ denotes the energy-momentum tensor of the field $\phi^{\mu}$ and $T_{\mu\nu}^{\text{S}}$ denotes the scalar contributions to the energy-momentum tensor. What we are considering here is a simplified STVG model without compromising essential conceptual difference with GR. The Kerr-MOG BH solution is based only on the vector field contributions to the field equations ignoring the scalar field contributions from $S_{\text{S}}$. Specifically, the Kerr-MOG BH was derived under the following assumptions \cite{PLR:2017}: (i) The mass of the vector field $\phi^{\mu}$ is neglected when $\widetilde{\mu}r << 1$ since the effect of $\widetilde{\mu}$ is manifest only at kiloparsec scales from the source \cite{MR:2014}. So it can be disregarded when solving the field equations for local compact objects such as a BH. (ii) $G$ is a constant that depends on the parameter $\alpha$:
\begin{equation}
G = G_{N}\left(1+\alpha \right) ,
\end{equation}%
where $G_{N}$ denotes Newton's gravitational constant, and $\alpha$ is a free adimensional STVG parameter.

Given these assumptions, the action (2) takes the simplified form
\begin{equation}
S = S_{\text{GR}} + S_{\phi} + S_{\text{M}} = \int d^{4}x\sqrt{-g}\left(\frac{R}{16\pi G} - \frac{1}{4}B^{\mu\nu}B_{\mu\nu}\right) + S_{\text{M}}.
\end{equation}%
Variation of the action with respect to $g_{\mu\nu}$ now yields the STVG field equations:
\begin{equation}
G_{\mu\nu} = 8\pi G\left(T_{\mu\nu}^{\phi} + T_{\mu\nu}^{\text{M}}\right),
\end{equation}%
where $G_{\mu\nu}$ is the Einstein tensor, and the energy-momentum tensor for the vector field $\phi^{\mu}$ is given by
\begin{equation}
T_{\mu\nu}^{\phi} = \frac{1}{4}\left(B_{\mu}^{\alpha}B_{\nu\alpha} - \frac{1}{4}g_{\mu\nu}B^{\alpha\beta}B_{\alpha\beta}\right).
\end{equation}%
Varying the action (12) with respect to the vector field $\phi^{\mu}$, we obtain the dynamical equation for such field as
\begin{equation}
\triangledown_{\nu}B^{\mu\nu} = J_{Q}^{\mu}
\end{equation}%
where Moffat \cite{Moffat:2015a} \textit{postulates} that the charge $Q$ of the vector field $\phi^{\mu}$ is of gravitational origin and is proportional to the BH mass $M$ (BHs are electric charge neutral):
\begin{equation}
Q = \pm\sqrt{\alpha G_{N}}M,
\end{equation}%
giving the matter 4-current density
\begin{equation}
J_{Q}^{\mu} = -\frac{2}{\sqrt{-g}}\frac{\delta S_{\text{M}}}{\delta \phi^{\mu}} = \sqrt{\alpha G_{N}}J_{\text{M}}^{\mu}.
\end{equation}

Under the assumed approximations, Eqs.(14) and (15) indeed resemble Einstein-Maxwell equations. However, the essential conceptual difference lies in the differences in nature of the sources for the vector fields $\phi^{\mu}$. In the Einstein-Maxwell theory, mass and electric charge currents are \textit{independent} properties of matter, and only the latter couples to the electromagnetic 4-vector field $A_{\mu}$. In STVG, on the other hand, every \textit{massive current} serves as a source that couples to the vector field $\phi^{\mu}$. This means that for a given matter distribution, $T_{\mu\nu}^{\text{M}}$ determines both the dynamics of the metric and of the vector fields. We see that in STVG theory the nature of the gravitational field has been modified with respect to GR in two ways: a change in the gravitational constant $G = G_{N}\left(1 + \alpha\right)$, and a vector field $\phi^{\mu}$ that exerts a gravitational Yukawa-type force, which we will show below.

To show the Yukawa-type force emerging from STVG, we consider the geodesic equation for a test particle in coordinates $x^{\mu}$ :($t$, $r$, $\theta$, $\varphi$) is given by
\begin{equation}
m\left(\frac{d^{2}x^{\mu}}{d\tau^{2}} + \Gamma_{\alpha\beta}^{\mu}\frac{dx^{\alpha}}{d\tau}\frac{dx^{\beta}}{d\tau}\right) = Q B_{\nu}^{\mu}\frac{dx^{\nu}}{d\tau}.
\end{equation}%
The radial equation of motion for material test particles in the weak field $r>>2GM$ and slow motion approximation $dx^{0}/d\tau \approx 1$, $B_{0}^{1} = d\phi _{0}/dr$ yields from Eq.(18)
\begin{equation}
m\left(\frac{d^{2}r}{dt^{2}} - \frac{J_{N}^{2}}{r^{3}} + \frac{GM}{r^{2}}\right) = Q\frac{d\phi_{0}}{dr},
\end{equation}%
where $J_{N}$ is the Newtonian orbital angular momentum. For the spherically symmetric static case, one finally obtains the radial acceleration \cite{PLR:2017}
\begin{equation}
a(r) = -\frac{G_{N}(1 + \alpha)M}{r^{2}} + G_{N}M\alpha\frac{\exp(-\widetilde{\mu}r)}{r^{2}}(1 + \widetilde{\mu}r).
\end{equation}%
This acceleration is the crucial feature of STVG - one sees the Yukawa-type repulsive contribution in the last term counterbalancing the enhanced Newtonian gravity, the first term, that is recovered in the weak field limit $\widetilde{\mu}r << 1$, or $r \rightarrow \infty$, as already assumed. Numerical values for $\alpha$ and $\widetilde{\mu}$ in the acceleration depend on the central BH mass $M$, exhibiting the scalar field nature of $G(x)$ and $\widetilde{\mu}(x)$. To date, no analytical functional solutions of such fields or \textit{extra degrees of freedom} have been proposed. The equations are very complicated and require numerical methods for solutions. However, numerical values for a wide range of central masses have been determined by adjusting the underlying phenomenology. For a review of such values, see Ref.~\refcite{MR:2013,BM:2007}.

The issue of extra degrees of freedom brings us to the \textit{no-hair conjecture}, that applies to scalar fields in GR (or the so-called no-scalar hair conjecture), is not challenged in the STVG sector where it resembles Einstein-Maxwell theory replaced only by massive vector Proca field $B_{\mu\nu}$. STVG is essentially a non-GR model with the vector field determined by matter current via Eq.(17). In the scalar field sector of $G(x)$ and $\widetilde{\mu}(x)$, for which there are no explicit known solutions yet, the status of the conjecture seems unclear and requires further study. Assuming that it holds, then the absence of such fields would mean $\alpha$ and $\widetilde{\mu}$ being identically zero, causing a return to GR in the large, and the STVG model simply falls. On the other hand, there exist many distinct non-GR testable predictions, where $\alpha$ and $\widetilde{\mu}$ are required to have non-zero values.

The metric of the Kerr-MOG black hole of mass $M$ and angular momentum $J=aM$ in STVG theory is \cite{Moffat:2015b}:
\begin{eqnarray}
ds^{2} &=&c^{2}\left(\Delta-a^{2}\sin^{2}\theta\right)\frac{dt^{2}}{\rho^{2}} + \frac{2ac\sin ^{2}{\theta }}{\rho ^{2}}\left[ (r^{2}+a^{2})-\Delta \right] dtd\phi \nonumber \\
&&- \frac{\rho^{2}}{\Delta}dr^{2} - \rho^{2}d\theta^{2} -\left[ (r^{2}+a^{2})^{2}-\Delta a^{2}\sin ^{2}\theta \right] \frac{\sin^{2}\theta }{\rho^{2}}d\phi^{2}, \\
\Delta &=&r^{2}-\frac{2GMr}{c^{2}}+a^{2}+\frac{\alpha G_{N}GM^{2}}{c^{4}} \nonumber \\
&=&r^{2}-\frac{2G_{N}(1+\alpha )Mr}{c^{2}}+a^{2}+\frac{\alpha (1+\alpha)G_{N}^{2}M^{2}}{c^{4}},  \nonumber \\
\rho ^{2} &=&r^{2}+a^{2}\cos^{2}\theta.  \nonumber
\end{eqnarray}%
The metric above reduces to the Kerr metric in GR when $\alpha =0$. By setting $a=0$ in Eq.(21), we recover the metric of a Schwarzschild STVG black hole.

The radius of the inner $r_{-}$ and outer $r_{+}$ event horizons are determined by the roots of $\Delta = 0$:
\begin{equation}
r_{\pm} = \frac{G_{N}(1 + \alpha)M}{c^{2}}\left[1 \pm \sqrt{1- \frac{a^{2}c^{4}}{G_{N}^{2}(1 + \alpha)M^{2}} - \frac{\alpha}{(1 + \alpha)}}\right].
\end{equation}
If we set $\alpha = 0$ in the latter expression, we obtain the formula for the inner and outer horizons of a Kerr black hole. Inspection of Eq.(22) also reveals that for $\alpha > 0$ the outer horizon of a Kerr black hole in STVG is larger that the corresponding one in GR.

The radial coordinate of the ergosphere is determined by the roots of $g_{tt} = 0$:
\begin{equation}
r_{E} = \frac{G_{N}M\left(1 + \alpha\right)}{c^{2}}\left[1 \pm \sqrt{1 - \frac{a^{2}\cos^{2}{\theta}c^{4}}{{G_{N}}^{2}\left(1 + \alpha\right)^{2}M^{2}} - \frac{\alpha}{1+\alpha}}\right].
\end{equation}%
We see that the ergosphere grows in size as the parameter $\alpha$ increases.

In the next section, we calculate and analyse the non-geodesic equatorial circular motion of test particles around these black holes. We develop the formalism for the calculation of the circular orbits in Kerr-STVG spacetime. The corresponding expressions in Schwarzschild-STVG spacetime can be easily obtained by setting $a=0$. By non-geodesic motion, we mean motion that is not in free fall but is acted on by forces other than gravity like, say, engine driven motion.

\section{Non-geodesic equatorial orbit}

Assume that the source/receiver, sending two oppositely directed light beams, is non-geodesically orbiting around a rotating black hole described by metric (21), along a circumference on the equatorial plane $\theta = \pi/2$. Such a motion of the source/receiver could be achieved by airborne clocks circumnavigating the Earth. Suitably placed mirrors send back to their origin both beams after a circular trip about the rotating central mass. Assume further that source/receiver is orbiting the central mass at a radius $r = R = $ const. Then the metric (21) reduces to
\begin{eqnarray}
d\tau^{2} &=& \left[1 - \frac{2(1+\alpha)M}{R} + \frac{\alpha(1+\alpha)M^{2}}{R^{2}}\right]dt^{2} + \frac{2a(1+\alpha)M}{R}\left(2-\frac{\alpha M}{R}\right) dtd\phi  \nonumber \\
&&-\left[R^{2}+a^{2}\left\{1 + \frac{2(1+\alpha)M}{R} - \frac{\alpha(1+\alpha)M^{2}}{R^{2}}\right\}\right] d\phi^{2}.
\end{eqnarray}%
The source/receiver is sharing the uniform spin $\omega_{0}$ of the Earth (non-geodesic motion, not determined by Kepler's law), the rotation angle $\phi_{0}$ of the source/receiver is
\begin{equation}
\phi_{0} = \omega_{0}t.
\end{equation}%
Then
\begin{equation}
d\tau^{2} = \left[1 - (a^{2}+R^{2})\omega_{0}^{2} - \left(2-\frac{\alpha M}{R}\right)\frac{(1+\alpha)M(1+a\omega_{0})^{2}}{R}\right]dt^{2}.
\end{equation}%
For light moving along the same circular path it must obey $d\tau = 0$. Assuming $\Omega$ to be the angular velocity of light motion along the path, we have
\begin{equation}
R^{2}\left\{1-\Omega^{2}\left(a^{2}+R^{2}\right)\right\} + (\alpha M-2R)(1+\alpha)M(1+a\Omega)^{2} = 0.
\end{equation}%
Solving Eq.(27), one finds two roots that represent the angular velocity $\Omega_{\pm}$ of light for the co- and counter rotating light motion given by
\begin{equation}
\Omega_{\pm} = \frac{a(1+\alpha )M(2R-\alpha M)\pm R^{2}\sqrt{R^{2}+a^{2} - 2(1+\alpha)MR + \alpha (1+\alpha )M^{2}}}{R^{4}+a^{2}\{R^{2}+2(1+\alpha)MR-\alpha (1+\alpha )M^{2}\}}.
\end{equation}

The rotation angles $\phi_{\pm}$ for light are then
\begin{equation}
\phi_{\pm} = \Omega_{\pm}t.
\end{equation}%
Eliminating $t$ between Eqs.(25) and (29), we obtain
\begin{equation}
\phi_{\pm} = \frac{\Omega_{\pm}}{\omega_{0}}\phi_{0}.
\end{equation}%
The first intersection of the world lines of the two light rays with the one of the orbiting source/receiver after the emission at time $t=0$ is, when the angles are
\begin{eqnarray}
\phi_{+}&=&\phi _{0}+2\pi , \\
\phi_{-}&=&\phi _{0}-2\pi ,
\end{eqnarray}
which give
\begin{equation}
\frac{\Omega_{\pm}}{\omega_{0}}\phi_{0} = \phi_{0}\pm 2\pi.
\end{equation}%
Solving for $\phi_{0}$,
\begin{equation}
\phi_{0\pm} = \mp\frac{2\pi\omega_{0}}{\Omega_{\pm} - \omega_{0}},
\end{equation}%
we have, putting the expressions for $\Omega_{\pm}$ from (28),
\begin{eqnarray}
\phi_{0\pm} &=& \mp 2\pi\omega_{0}/\left[-\omega_{0} + \left\{a(1+\alpha)M(2R-\alpha M)\pm R^{2}\left(R^{2}+a^{2} - 2(1+\alpha)MR \right.\right.\right. \nonumber\\
&&\left.\left.\left. + \alpha (1+\alpha)M^{2}\right)^{1/2}\right\}/\left\{R^{4}+a^{2}7\left(R^{2}+2(1+\alpha)MR-\alpha (1+\alpha )M^{2}\right)\right\}\right].
\end{eqnarray}

The proper time of the rotating source/receiver, deduced from Eq.(26) using Eq.(25), is
\begin{equation}
d\tau = \left[1 - \left(1 + \frac{R^{2}}{a^{2}}\right)a^{2}\omega _{0}^{2} - \left( 2-\frac{\alpha M}{R}\right) \frac{(1+\alpha )M(1+a\omega _{0})^{2}}{R}\right] ^{\frac{1}{2}}\frac{d\phi _{0}}{\omega _{0}}.
\end{equation}%
Finally, integrating between $\phi_{0-}$ and $\phi_{0+}$, we obtain the \textit{exact} Sagnac delay
\begin{equation}
d\tau = \left[1 - \left(1 + \frac{R^{2}}{a^{2}}\right)a^{2}\omega _{0}^{2} - \left( 2-\frac{\alpha M}{R}\right) \frac{(1+\alpha )M(1+a\omega _{0})^{2}}{R}\right] ^{\frac{1}{2}}\frac{\phi _{0+}-\phi _{0-}}{\omega _{0}}.
\end{equation}%

Using the integration limits from Eq.(35), we explicitly write it out as
\begin{eqnarray}
\delta \tau &=& \frac{4\pi}{R}\left[\left\{R(R^{2}+a^{2})+2a^{2}(1+\alpha)M-\frac{a^{2}\alpha(1+\alpha)M^{2}}{R}\right\}\omega_{0}\right. \nonumber \\
&& \left.-a(1+\alpha)M\left(2- \frac{\alpha M}{R}\right)\right]/ \left[1-\frac{(1+\alpha)M(1-2a\omega_{0})}{R}\left(2-\frac{\alpha M}{R}\right)\right.  \nonumber \\
&&\left.- \left\{R^{2}+a^{2}+\frac{a^{2}(1+\alpha)M}{R}\left(2-\frac{\alpha M}{R}\right)\right\}\omega_{0}^{2}\right]^{\frac{1}{2}}.
\end{eqnarray}%
Then expanding, we find
\begin{eqnarray}
\delta\tau &=& \frac{4\pi\left[\left\{R(R^{2}+a^{2})+2a^{2}M\right\}\omega_{0}-2aM\right]}{R\sqrt{1-2M(1-2a\omega_{0})/R-\{R^{2}+a^{2} + 2a^{2}M/R\}\omega_{0}^{2}}} \nonumber\\
&& +\left[2\pi M(2R-M)(1-a\omega_{0})\left\{2a(M-R) \right.\right. \nonumber \\
&&\left.\left.+(R^{3}-4Ma^{2}+a^{2}R)\omega_{0}+(R^{3}+2Ma^{2}+a^{2}R)a\omega_{0}^{2}\right\}\right]\alpha / \nonumber \\
&&[R\{R-2M+2aM\omega_{0}-(R^{3}+a^{2}R+2Ma^{2})\omega_{0}^{2}\}]^{\frac{3}{2}} + O(\alpha^2).
\end{eqnarray}

Sagnac delay (38) is often loosely interpreted as the gravitational analogue of the Bohm-Aharonov effect \cite{AB:1959} although it is not quite accurate. The best analogue of the Bohm-Aharonov effect could be when light beams move along a flat space torus (see for details, Ref.~\refcite{Semon:1982}). Nevertheless, as shown by Ruggiero \cite{Ruggiero:2004}, expression (38) completely agrees with the one of the gravito-electromagnetic Bohm-Aharonov interpretation \cite{Sakurai:1980}. For the viewpoint of Bohm-Aharonov quantum interference in GR, see Ref.~\refcite{NZ:2002,AEN:2001}.

We can imagine a source/receiver keeping a fixed position in a coordinate system defined by distant fixed stars ($\omega_{0} = 0$). A Sagnac delay for such clocks will occur under the condition that $a \neq 0$, given by
\begin{equation}
\delta\tau_{0} = -\frac{4\pi a(1+\alpha)M(2R-\alpha M)}{R\sqrt{R^{2}-2(1+\alpha)MR + \alpha(1+\alpha)M^{2}}}.
\end{equation}%
Interestingly, not the shape but the closedness of the orbit is the main thing causing the delay. This has been shown by Cohen and Mashhoon \cite{CM:1993} sending a pair of light beams in opposite directions along a closed triangular circuit, instead of a circle. They found the same result as above in that approximation. Eq.(38) is the final result for the Sagnac delay for the equatorial non-geodesic motion. In most cases many terms in this equation are very small allowing series approximations, which we do below. Let us first assume that $\beta = \omega_{0}R\ll 1$, and develop Eq.(38) in powers of $\beta$ retaining terms only up to the second order. The result is
\begin{eqnarray}
\delta \tau  &\simeq &-\frac{4\pi a(1+\alpha )M(2R-\alpha M)}{R\sqrt{R^{2}-2(1+\alpha )MR+\alpha (1+\alpha )M^{2}}} \nonumber \\
&&+ \frac{4\pi R^{2}\left\{R^{2}-2(1+\alpha )MR+a^{2}+\alpha (1+\alpha )M^{2}\right\}}{\left\{R^{2}-2(1+\alpha )MR+\alpha (1+\alpha )M^{2}\right\}^{3/2}}\beta,
\end{eqnarray}%
which displays that the first term is just $\delta\tau_{0}$ of Eq.(40), as expected. Now we perform a successive post-Newtonian approximation in $\varepsilon = M/R\ll 1$ and in $a/R\ll 1$, and using the expression $\delta\tau_{S} = 4\pi\beta R$, we obtain by expansion
\begin{equation}
\delta\tau \simeq \delta\tau_{S} - 8\pi a(1+\alpha )\varepsilon + 4\pi R(1+\alpha)\varepsilon\beta .
\end{equation}

Restoring $\varepsilon$ and $\beta$ we obtain
\begin{equation}
\delta\tau\simeq\delta \tau_{S} -\frac{8\pi a(1+\alpha)M}{R} + 4\pi R(1+\alpha)M\omega_{0}.
\end{equation}
The effect of $\alpha$ is evident.

\section{Geodesic equatorial orbit}

We are considering a circular geodesic orbit of the source/receiver (maybe a free fall satellite) at some arbitrary radius on the equator ($\theta = \pi/2$) and sending light signals circumnavigating the Earth. The Lagrangian $\mathcal{L}$ for geodesic motion is given by
\begin{equation}
\mathcal{L} = \frac{1}{2}g_{\mu\nu}\dot{x}^{\mu}\dot{x}^{\nu}
\end{equation}%
and the Euler-Lagrange $r-$equation is
\begin{equation}
\frac{d}{d\tau }\left( \frac{\partial \mathcal{L}}{\partial \dot{r}}\right) =%
\frac{\partial \mathcal{L}}{\partial r}.
\end{equation}%
Since in metric (21), $g_{r\mu}=0$ for $r\neq\mu$, we have%
\begin{equation}
\frac{d}{d\tau}\left(g_{rr}\dot{r}\right) = \frac{1}{2}g_{\mu\nu,r}\dot{x}^{\mu}\dot{x}^{\nu}.
\end{equation}%
Circular orbits are defined by
\begin{equation}
\dot{r} = \ddot{r} = 0,
\end{equation}%
and the Eq.(46) yields%
\begin{equation}
g_{tt,r}\dot{t}^{2} + 2g_{t\phi,r}\dot{t}\dot{\phi} + g_{\phi\phi,r}\dot{\phi}^{2} = 0.
\end{equation}%
Defining $\omega = \dot{\phi}/\dot{t}$, this equation yields the quadratic equation
\begin{equation}
g_{\phi \phi ,r}\omega ^{2}+2g_{t\phi ,r}\omega +g_{tt,r}=0.
\end{equation}

From the metric (21), putting $dr = 0$ at $r = R = $ const. and $d\theta = 0$ at $\theta = \pi/2$, we find
\begin{equation}
d\tau^{2} = g_{tt}dt^{2} + 2g_{t\phi}dtd\phi + g_{\phi\phi}d\phi^{2},
\end{equation}%
where%
\begin{eqnarray}
g_{tt}       &=& 1 - \frac{2(1+\alpha)M}{R} + \frac{\alpha(1+\alpha)M^{2}}{R^{2}}, \\
g_{t\phi}    &=& \frac{a(1+\alpha)M(2R-\alpha M)}{R^{2}}, \\
g_{\phi\phi} &=& - R^{2} - a^{2}\left\{1 + \frac{2(1+\alpha)M}{R} - \frac{\alpha(1+\alpha)M^{2}}{R^{2}}\right\}.
\end{eqnarray}%

The source/receiver rotational velocities $\omega_{\pm}$ then follow from the two roots of Eq.(49), and using Eqs.(51)-(53) in them, we obtain
\begin{equation}
\omega_{\pm} = \frac{a(1+\alpha)M(R-\alpha M)\pm R^{2}\sqrt{(1+\alpha)M(R-\alpha M)}}{a^{2}(1+\alpha)M(R-\alpha M)-R^{4}}.
\end{equation}

The exact Sagnac delay for geodesic motion then is $\delta\tau_{S\pm\text{geo}}^{\text{Kerr STVG}} = 4\pi\omega_{\pm}R^{2}$, which can be written out explicity as below:
\begin{eqnarray}
\delta\tau_{S\pm\text{geo}}^{\text{Kerr STVG}} &=& \pm 4\pi \left[\frac{\sqrt{MR^{7}}\pm aMR^{2}}{R^{3}-a^{2}M} +\frac{\sqrt{MR^{5}}(R-M)(R^{3}+a^{2}M)}{2(R^{3}-a^{2}M)^{2}}\alpha \right. \nonumber \\
&&\left.\pm\frac{2aMR^{4}(R-M)}{2(R^{3}-a^{2}M)^{2}}\alpha+O(\alpha^{2})\right].
\end{eqnarray}%

Under the weak field conditions $\frac{a^{2}}{R^{2}}\ll 1$, $\frac{M}{R}\ll 1$, the first two expressions below then coincide with the formula derived by \cite{LI:2011,Karimovetal:2018a}:
\begin{eqnarray}
\delta \tau _{\pm \text{geo}}^{\text{KerrSTVG}}&=&\pm 4\pi \sqrt{MR}\mp 4\pi
\left( \frac{M}{R}\right) a \nonumber \\
&&\pm 4\pi \left[\frac{\sqrt{MR^{5}}(R-M)(R^{3}+a^{2}M)}{2(R^{3}-a^{2}M)^{2}} \right.\left.\pm\frac{2aMR^{4}(R-M)}{2(R^{3}-a^{2}M)^{2}}\right] \alpha.
\end{eqnarray}%
The Kerr terms follow from the expansion of (56) at $\alpha =0$. With $a=0$, $\alpha =0$, we get the Schwarzschild value
\begin{equation}
\delta \tau _{S\pm \text{geo}}^{\text{Sch}}=\pm 4\pi \sqrt{MR}.
\end{equation}

\section{Upper limits from the terrestrial Sagnac delay}

The earliest experiment measuring terrestrial Sagnac delay or synchrony gap was conducted by around-the-Earth airborne clock experiment by Hafele and Keating \cite{HK:1972}. A more precise later experiment was conducted by Allan, Weiss and Ashby (AWA)\cite{AWA:1985}. Instead of portable clocks, they used four GPS satellites transmitting electromagnetic signals that can have a common view from remote stations on Earth. The experiment is equivalent to $90$ day of independent Hafele-Keating runs yielding flat space one-way delays from about $240$ to $350$ ns with Sagnac error residual of only $5$ ns. (One way delay is $\frac{1}{2} \delta\tau_{S}$, which means that there is a stationary clock on Earth, and its reading is compared with that of the airborne clock after its non-geodesic circumnavigation). Any of the measured values of $\frac{1}{2} \delta\tau_{S}$ can be used in the expression $\delta\tau_{S} = \frac{4\omega {0} S}{c^{2}}$, where $S$ is the area of the projection, orthogonal to the rotation axis, of the closed path followed by the waves contouring the turntable, $c$ is the speed of light in vacuum and $\omega_{0}$ is the angular velocity of the turntable, to obtain the desired equivalent radius. Allan, Weiss and Ashby  \cite{AWA:1985} used GPS satellites around Earth moving with Earth's angular speed $\omega_{0}$ used by the Hafele-Keating non-geodesic motion (58).

The relevant Earth data are as follows:
\begin{eqnarray}
\omega_{0} &=& \Omega_{\oplus} = 7.303\times 10^{-5} \;\text{rad/s} \Rightarrow \nonumber \\
2\omega _{0}/c^{2} &=& 1.622\times 10^{-21} \;\text{rad}\left(\text{s/m}^{2}\right), \\
M&\rightarrow& GM_{\oplus }/c^{2} = 4.35\times 10^{-3} \;\text{\textmd{m}}, \\
a &=& a_{\oplus} = \left(\frac{2}{5}\right)R_{\oplus}^{2}\omega_{0} = 1.18\times 10^{9} \;\text{m}^{2}\text{/s}, \\
c &=& 3\times 10^{8} \;\text{\textmd{m/s}}.
\end{eqnarray}%
where $a$ is the specific surface angular momentum. The Hafele and Keating \cite{HK:1972} delay is given by the formula $\delta\tau_{S} = 4\pi\omega_{0}R^{2}/c^{2}$. With $\omega_{0} = \Omega_{\oplus}$ and the \textit{average radius} of their circumnavigating equatorial flight $R = R_{\oplus} = 6378137$ m, due to the east and westward equatorial motion of the airborne atomic clocks, it works out to
\begin{equation}
\delta \tau _{S}=2\times \frac{2\Omega _{\oplus }}{c^{2}}\times \pi
R_{\oplus }^{2}=4.148\times 10^{-7}\;\text{\textmd{s}}=2\times 207.4\;\text{%
\textmd{ns}}.
\end{equation}%
As well known, this is the famous value $\frac{1}{2}\delta\tau_{S}(=207.4$ ns) of the measured one way delay with error residual $\sim 10$ ns.

Allan, Weiss and Ashby \cite{AWA:1985} measured a range of values for the Sagnac delay between $240$ to $350$ ns. For fixing values, let us take $\frac{1}{2}\delta\tau_{S} = 240$ ns, which then yields an \textit{equivalent} Earth radius $R_{\text{eq}} = 6.86098\times 10^{6}$ m corresponding to a conventional circumnavigation. Putting $R = R_{\text{eq}}$ and the above Earth data (58)-(61) into $\frac{1}{2}\delta\tau$ of Eq.(43), restoring $\omega_{0} \rightarrow \Omega_{\oplus}/c^{2}$, $a \rightarrow a_{\oplus}/c^{2}$, and converting second to nanosecond (ns) scale, $\frac{1}{2}\delta\tau \rightarrow \frac{1}{2}\delta\tau \times 10^{9}$ ns, we find the values of one-way individual terms after converting to nanosecond by multiplying each pieces by $10^{-9}$ as below%
\begin{eqnarray}
\delta\tau &\simeq& \delta\tau_{S}^{\text{AWA}} - \frac{8\pi a(1+\alpha)M}{R} + 4\pi R(1+\alpha)M\omega_{0}, \nonumber \\
\left(\frac{1}{2}\right)\delta\tau_{S}^{\text{AWA}} &=& 2\times\frac{2\Omega_{\oplus}}{c^{2}} \times \pi R_{\text{eq}}^{2}\times 10^{9} = 240\; \text{ns}, \nonumber \\
\left(\frac{1}{2}\right)\delta\tau_{a_{\oplus}} &=& \left(\frac{1}{2}\right)\frac{8\pi aM}{R} = \left(\frac{1}{2}\right) \frac{8\pi a_{\oplus}M}{c^{2}R_{\text{eq}}}  \nonumber \\
&\rightarrow& \frac{1}{2}\delta \tau _{M, a_{\oplus}}\times 10^{9}=1.04\times 10^{-7}\text{ ns}, \\
\left(\frac{1}{2}\right)\frac{8\pi\alpha aM}{R} &=& \left(\frac{1}{2}\right)\frac{8\pi\alpha a_{\oplus}M}{c^{2}R_{\text{eq}}}  \nonumber \\
&\rightarrow& \frac{1}{2}\delta\tau_{\alpha, M, a_{\oplus}}\times 10^{9} = 1.04\times 10^{-7}\alpha \text{ ns}, \\
\left(\frac{1}{2}\right) 4\pi RM\omega_{0} &=& \left(\frac{1}{2}\right)\frac{4\pi R_{\text{eq}}M\Omega_{\oplus}}{c^{2}}  \nonumber \\
&\rightarrow& \frac{1}{2}\delta\tau_{M, \Omega_{\oplus}}\times 10^{9} = 1.52\times 10^{-14} \text{ ns}, \\
\left(\frac{1}{2}\right) 4\pi RM\omega_{0} &=& \left(\frac{1}{2}\right)\frac{4\pi R_{\text{eq}}M\Omega_{\oplus}\alpha}{c^{2}}  \nonumber \\
&\rightarrow& \frac{1}{2}\delta\tau_{\alpha, M, \Omega_{\oplus}}\times 10^{9} = 1.52\times 10^{-14}\alpha \text{ ns},  \nonumber
\end{eqnarray}

The contributions $1.04\times 10^{-7}$ ns and $1.52\times 10^{-25}$ ns are already far less that the observed error residual of $5$ ns. Hence, to leading order, one essentially has
\begin{equation}
\frac{1}{2}\delta\tau\times10^{9} \simeq \frac{1}{2}\delta\tau_{S}^{\text{AWA}} + 1.04\times 10^{-7}\alpha.
\end{equation}%
Since the Earth values of $M$ and $a$ are already plugged in Eq.(43), only $\alpha$ is appearing in (63). Using the assumption, as in Refs.\refcite{Kulbakovaetal:2018,Karimovetal:2018b}, that the total one way correction term should be less than or equal to the average error residual of $5$ ns, we may obtain an upper bound on $\alpha$ as
\begin{equation}
1.04\times 10^{-7}\alpha \leq 5 \Rightarrow \alpha \leq 4.80\times 10^{7}.
\end{equation}%
We shall see in the next section that the upper bound can be considerably refined using updated data.

\section{Global Navigation Satellite Systems update}

Updated data on clock synchronization using two-way radio links between two ground stations and a freely falling (geodesic motion) satellite are available. The satellite systems are equipped with accurate, stable atomic clocks on-board, while there are precision clocks fixed on the ground providing world-wide access to position, velocity and time of all events. An excellent recent review by Ashby \cite{Ashby:2014} enumerates the various relativistic factors that have to be accounted for if the systems have to work well. These factors include relativistic principles, concepts and effects such as the constancy of the speed of light, relativity of synchronization, coordinate time, proper time, time dilation, the Sagnac effect, the weak equivalence principle and gravitational frequency shifts. Additionally, Shapiro time delay and tidal effects caused by the moon and the sun might also be corrected for in the future experiments. See Ref. \refcite{Ashby:2003} for details.

The GPS satellites circle the Earth at an orbital radius of 26,000 km, corresponding to a 12 hour orbit. They complete two full orbits every day, but for the Sagnac delay to appear, one full orbit is enough. The GPS satellites are not in a geostationary orbit, but rise and set two times per day. Each satellite broadcasts radio waves towards Earth that contain information regarding its position and time. We shall use the updated \textit{fluctuation} in the delay as observed in the GPS clock system after circumnavigation along orbital radius $R = R_{\text{geo}} = 4.076 R_{\oplus} = 2.6\times 10^{7}$ m. The fluctuation is caused by the jittery motion of the satellite about the equator. When the positions of the Earth stations are fixed and the satellite moves strictly over the equator (at $R_{\oplus} = 6378137$ m), the observed Sagnac delay ($207.4$ ns) is constant. In reality, however, the actual position of the orbiting satellite at an orbital radius $R = R_{\text{geo}}$ varies slightly over a $12-$hr period. A study demonstrated that the time varying fluctuation in the delay due to satellite jitter could add more than $0.5$ ns to the constant basic value \cite{Fujiedaetal:2006}.

In the original Hafele-Keating experiment, clocks are transported round the Earth by engine-driven aircrafts \cite{HK:1972} at $R = R_{\oplus}$ along the equator \cite{AWA:1985}. On the other hand, after initial launching, GPS\ clocks should necessarily follow a force-free geodesic orbit occurring at the radius $R = R_{\text{geo}} = 2.6 \times 10^{7}$ m. The non-trivial changes in the two-way Sagnac delay due to the change from non-geodesic to geodesic orbit can in fact be quite large:
\begin{eqnarray}
\delta\tau_{S\;\text{nongeo}}^{\text{HK}} &=& \frac{4\pi\Omega_{\oplus}R_{\oplus}^{2}}{c^{2}} = 414.8 \;\text{\textmd{ns},}
\;\left(\omega_{0} = \Omega_{\oplus}, \;R = R_{\oplus}\right), \\
\delta\tau_{S\;\text{geo}}^{\text{Sch}} &=& \frac{4\pi\Omega_{\oplus}R_{\text{geo}}^{2}}{c^{2}} = 6893.1 \;\text{\textmd{ns},}
\;\left(\omega_{0} = \Omega_{\oplus}, \;R = R_{\text{geo}}\right),
\end{eqnarray}%
where $\Omega_{\oplus} = 7.303\times 10^{-5}$ rad/s has been used. This shows that the delay from the geodesic motion (51) would be $16.61$ times larger than that from the non-geodesic motion. The force-balance equation for geodesic orbits gives%
\begin{equation}
\Omega_{\oplus} = \sqrt{GM_{\oplus}/R_{\text{geo}}^{3}}
\end{equation}%
which yields, restoring $c$,
\begin{equation}
\delta\tau_{S\;\text{geo}}^{\text{Sch}} = \left(\frac{4\pi}{c}\right)\sqrt{\left(GM_{\oplus}/c^{2}\right)R_{\text{geo}}}.
\end{equation}%
This is just the first term or the Schwarzschild expression (57) for the delay. This term will appear in the leading order expansion of exact GR master expression for the Sagnac delay (56) for geodesic motion in the STVG metric. Letting the parameters assume the Earth values $a = a_{\oplus}$, $M=M_{\oplus}$ and the geodesic orbit $R = R_{\text{geo}}$ in Eq.(56), and converting second to nanosecond as before, we find (taking only the positive sign inside the square bracket,
\begin{eqnarray}
\left\vert\delta\tau_{S\;\text{geo}}^{\text{Kerr STVG}}\right\vert &=& \delta\tau_{S\;\text{geo}}^{\text{Sch}} + \text{correction due to spin}\;a_{\oplus} +\text{correction due to}\;\alpha  \nonumber \\
&=& \left[1.40\times 10^{4} + 8.27 + 7.05\times 10^{3}\alpha\right]\;\text{\textmd{ns}}.
\end{eqnarray}

Like before, our idea is to identify the above $\alpha-$correction with the fluctuation in the observed updated data on the round-trip Sagnac delay. In this context, we note that precision measurement of the Two-Way Satellite Time and Frequency Transfer (TWSTFT) highly depends on the residual non-reciprocity delays -- one of them is caused by the Sagnac delay or synchronization discontinuity found between the re-united flying clocks \cite{Schlegel:1973}. GPS calibration of clocks by means of TWSTFT involving Sagnac delay has achieved an accuracy at the level of nanoseconds \cite{Fujiedaetal:2006,Piesteretal:2008}.

Tseng {\it et al.} \cite{Tsengetal:2011} showed that the Sagnac delay can be calculated using the Earth station coordinates and the actual ephemeris data. Their experiment involved Earth stations located at the National Institute of Information and Communications Technology (NICT) in Japan and the Telecommunication Laboratories (TL) in Taiwan. This experiment on the time rate of variation of the Sagnac delay, called diurnal \cite{Tsengetal:2011}, predicted a variation $\Delta(\delta\tau_{{S}})$ of magnitude $\pm 0.25$ ns, which provides an update over $\sim 5$ ns obtained in the Allan, Weiss and Ashby \cite{AWA:1985}. This updated value from TWSTFT then provides an upper bound on the correction term in Eq.(67). Looking at Eq.(72), we see that the Sagnac contribution due to Earth's spin $a_{\oplus}$ adds a little, $8.27$ ns, to the basic $\delta\tau_{S\;\text{geo}}^{\text{Sch}}$ but is already far bigger than the diurnal $\pm 0.25$ ns. This contribution does not involve the STVG parameter $\alpha$. The last term in Eq.(72) involves $\alpha$, and using the same assumption, as in Ref. \refcite{Kulbakovaetal:2018,Karimovetal:2018b}, that the leading order correction term due to $\alpha$ should also be less than or equal to the fluctuation $\Delta(\delta\tau_{{S}})$ ns, so we have
\begin{equation}
\alpha < \frac{\Delta(\delta\tau_{{S}})}{7.05\times 10^{3}} = \frac{0.25}{7.05\times 10^{3}} = 3.54\times 10^{-5},
\end{equation}%
which constrains $\alpha$ much more severely than the one obtained in (67).

\section{Conclusions}

Observations of rotation curves of galaxies, mass profiles of galaxy clusters, cosmological data require the postulate of an invisible dark matter within the framework of GR. However, every experiment to date has failed to measure the properties of such dark matter. A natural alternative then is to look for gravity theories that do not require the postulate of dark matter/energy components. In this quest, Moffat \cite{Moffat:2006} developed his STVG theory that can explain not only the Solar system tests, but also the galaxy rotation curves \cite{MR:2013}, clusters dynamics \cite{MR:2014}, Bullet Cluster phenomena \cite{BM:2007}, and cosmological data \cite{MT:2013} without requiring the \textit{undetected} GR dark components of matter. Further, recent astronomical determinations of neutron star masses \cite{Demorestetal:2012,Ozeletal:2012,Antoniadisetal:2013,Kiziltanetal:2013} defy GR limits, whereas there is good agreement with the predictions from the STVG \cite{LR:2017}. All these indicate that STVG could be a potential candidate for a new gravity theory. In the STVG, the gravitational coupling constant $G$ is assumed to exceed Newton's constant $G_{N}$ as $G = G_{N} (1 + \alpha)$, $\alpha > 0$. This assumption together with introduction of a Yukawa-like gravitational acceleration is mostly responsible for the successes of the theory. Therefore, it is necessary to investigate what constraints appear on the STVG parameter $\alpha$ from different physical considerations. In the literature, it is constrained to be in the range $0.03 < \alpha < 2.47$ corresponding to different central source masses and accretion characteristics have been studied for different values of $\alpha$ \cite{PLR:2017}.

In this paper, we attempted to constrain the range of $\alpha$ even further. For this purpose, we considered the terrestrial Sagnac experiment and \textit{assumed that the $\alpha$-dependent leading order contribution~to the delay should not exceed the error residual or fluctuation due to jittery motion of the satellite in the delay measurement}. The non-geodesic motion of clocks in the earlier Hafele and Keating \cite{HK:1972} run once and in the later precision experiment by Allan, Weiss and Ashby \cite{AWA:1985} yielded an error residual of $\sim 5$ ns that led to $\alpha < 4.8 \times 10^{7}$. This is rather a large upper bound probably not much informative.

Hence, we considered the updated fluctuation in the GPS measurement of the terrestrial Sagnac delay that uses geodesic force-balance equation. The fluctuation is caused by the jittery motion of the satellite about the equator. When the positions of the Earth stations are fixed and the satellite moves strictly over the equator, the basic Sagnac delay is constant. In reality, however, the actual position of the geodesically moving satellite varies slightly over a $12-$hr period. The experiment (see Ref. \refcite{Tsengetal:2011}) on the time rate of variation of the Sagnac delay predicted a fluctuation $\Delta (\delta \tau _{{S}})$ of magnitude $\pm 0.25$ ns, which provides an update over the $\sim 5$ ns error. Under the assumption adopted in this paper, it led to a constraint $0<\alpha <10^{-5}$ corresponding to Earth's mass, which suggests a further improvement on the upper bound $\alpha <0.1$ proposed by Lopez Armengol and Romero \cite{LR:2017} for stellar mass sources in a different experiment.

\section*{Acknowledgments}

We are deeply indebted to an anonymous reviewer for suggesting several improvements. The reported study was funded by RFBR according to the research project No. 18-32-00377.

%\begin{thebibliography}{000} %for 3 digits
%\begin{thebibliography}{00}  %for 2 digits

\end{document}